# Capítulo 2

# Una breve historia social de la astrobiología en Iberoamérica

Guillermo A. Lemarchand

**Resumen.** El trabajo se divide en tres secciones: en la primera se describe la evolución histórica de los principales argumentos presentados acerca de la pluralidad de mundos habitados, de los presocráticos al nacimiento de la ciencia moderna. En la segunda se analiza la puja por delimitar la actividad científica de búsqueda de vida fuera de la Tierra bajo una denominación específica. Finalmente, en la tercera parte, se presenta una breve descripción de la historia social de la ciencia, que permitió el incipiente desarrollo de la astrobiología en Iberoamérica.

**Abstract**. The work is divided into three sections: the first one describes the historical evolution of the main arguments presented about the plurality of inhabited worlds, from the presocratics to the birth of modern science. The second section analyzes the race to define the search for life beyond Earth as a scientific activity under a specific name. Finally, the third part presents a brief description of the social history of science that allowed the early development of astrobiology in Iberoamerica.

Guillermo A. Lemarchand (✉)

Consultor Regional del Programa de Ciencias Básicas e Ingeniería (2008-2010) de la Oficina Regional de Ciencia de la UNESCO para América Latina y el Caribe y Director del Proyecto SETI en el Instituto Argentino de Radioastronomía, C.C. 8 –Sucursal 25, C1425ZAB, Buenos Aires, Argentina

`lemar@correo.uba.ar`

## 1. Introducción al concepto de pluralidad de mundos habitados: del paradigma aristotélico al nacimiento de la ciencia moderna

Las raíces del pensamiento acerca de la pluralidad de mundos habitados, se remonta a la antigua Grecia. Uno de sus más destacados defensores fue Epicuro (341-270 a.C.), quien desarrolló ciertas ideas basadas en Leucipo (siglo V a. C) y Demócrito (460 - 370 a. C.) donde postulaba que: (1) la materia está compuesta por átomos, (2) el presente estado de la naturaleza es resultado de un largo proceso evolutivo y (3) la vida existe en todas partes en el Universo. En una carta que Epicuro le escribiera a Herodoto (Bailey 1957) se puede encontrar el siguiente fragmento que demuestra la convicción pluralista de esta escuela de pensamiento:

*"...hay infinitos mundos como el nuestro y distintos al él. Los átomos que son infinitos en número… nacen en las profundidades del espacio. Debemos asumir que un número finito de átomos forman los mundos como el nuestro, también aquellos mundos distintos al nuestro y aquellos que no pertenecen a ninguna de estas categorías. Por lo tanto, no existe ningún obstáculo para la existencia de un número infinito de otros mundos…. Debemos creer que en todos los mundos hay criaturas vivientes, plantas y otras cosas que encontramos en nuestro propio mundo".*

Siglos más tarde, el historiador Plutarco (46-120 d.C.) desarrolló el primer esbozo de lo que actualmente se define como *Principio de Mediocridad* (Hoerner 1961). Planteó la tesis de la pluralidad de mundos habitados basado en las siguientes cuatro premisas: (1) la Tierra no ocupa ninguna posición privilegiada en el universo; (2) la Tierra y los cuerpos pesados no constituyen el centro del universo según postula el paradigma aristotélico y están distribuidos homogéneamente por una mente superior; (3) la Luna es lo suficientemente parecida a la Tierra como para sustentar vida y (4) si no existiera vida en la Luna no tendría sentido su propia existencia.

Se debe destacar que todas estas especulaciones filosóficas tuvieron lugar en un ambiente en donde el paradigma aristotélico era hegemónico y donde la sola idea de existencia de otros mundos similares a la Tierra contradecía sus propios cimientos. Para Aristóteles (384-322 a.C.) y sus seguidores el *elemento tierra* ocupaba el centro del universo y era seguido secuencialmente por los elementos *agua*, *aire, fuego* y *éter*. En este paradigma, cada elemento tenía la naturaleza de moverse siempre en dirección de su esfera de pertenencia. Por esta razón, simplemente no era posible encontrar el elemento tierra en los cielos, ya que de existir debería manifestar movimientos violentos en dirección al centro del universo (la Tierra) y este hecho no era observado. Por lo tanto, para los aristotélicos era imposible concebir la sola existencia de otros mundos habitados.

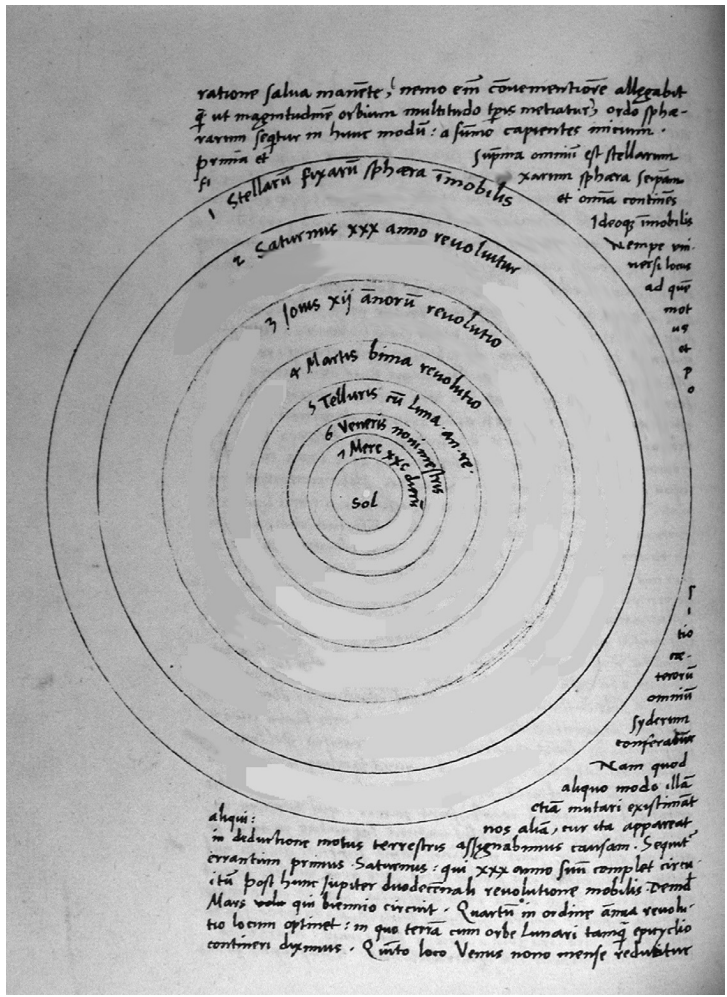

**Fig. 1.** Manuscrito de Nicolás Copérnico en donde plantea el primer modelo heliocéntrico del sistema solar. Foto: Guillermo A. Lemarchand tomada en la Colección de Libros Raros de la Universidad de Colorado en Boulder.

Con el advenimiento del nuevo paradigma copernicano, nuevas ideas acerca de la pluralidad de mundos habitados comenzaron a surgir. Sin duda uno de sus más audaces defensores fue Giordano Bruno (1548-1600), cuya pasión por los nuevos y osados conocimientos era escasamente más limitada que la infinitud del universo que reclamaba. En 1584 publica *Del infinito universo y sus mundos*, texto del cual se extrajo el siguiente diálogo:

> *Burquio: ¿Así pues que los otros mundos están habitados como este?*
>
> *Fracastorio: Si no es así y de mejor modo, por lo menos igualmente, porque es imposible que un espíritu racional y un tanto despierto pueda imaginar que carezcan de parecidos y mejores habitantes innumerables mundos que se revelan tan magníficos, o más que este, los cuales o son soles o no reciben menos que el Sol, los divinísimos y fecundos rayos que tanto nos revelan la felicidad de su propio sujeto y fuente como hacen dichosos a los circunstantes que participan de tal fuerza difundida. Son pues, infinitos los innumerables y principales miembros del universo, que tienen igual rostro, aspecto, prerrogativas, fuerzas y efecto.*

El 17 de febrero de 1600, en Roma, la Inquisición trasladó a Bruno a su lugar de ejecución. Su crimen: herejía. Entre sus creencias figuraban que la Tierra no era el centro del universo, había un infinito número de mundos y la vida existía en ellos… La sentencia: debería ser quemado en la hoguera.

Ante tan horroroso espectáculo, Galileo Galilei (1584-1642), se manifestó más prudente y en su *"Carta sobre las manchas solares"* escribió lo siguiente:

> *"Es posible creer en la probabilidad de que existan seres vivientes y vegetales en la Luna y los planetas, cuyas características no solo los hagan distintos a los terrestres, sino sumamente diferentes de aquellos que puede imaginar nuestra salvaje imaginación. Por mi parte, no puedo afirmar o negar dicha posibilidad. Dejo esta decisión a hombres más eruditos que yo."*

En 1610, Johanes Kepler (1571-1630) redacta un pequeño texto en respuesta a los sorprendentes descubrimientos que Galileo había realizado con su telescopio, dentro de sus páginas se encuentra el siguiente párrafo:

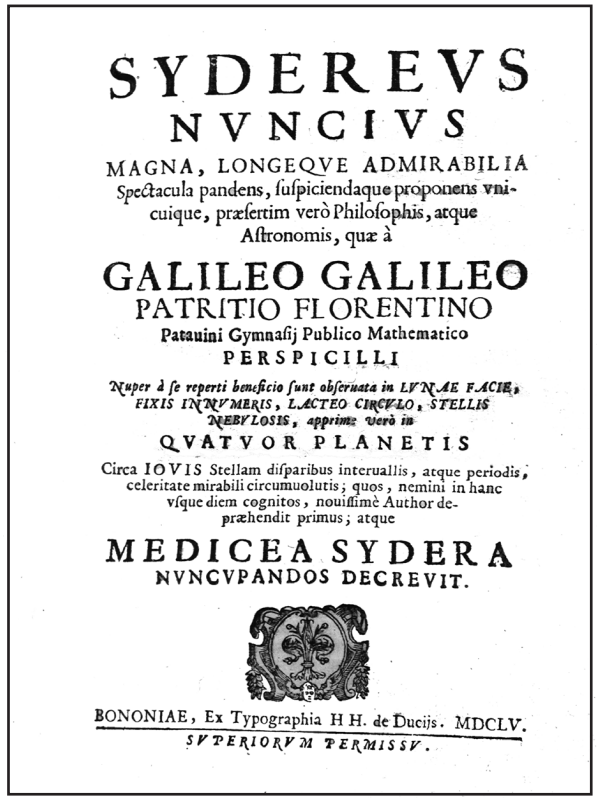

**Fig.2.** Carátula del libro *"El Mensajero Sideral"* escrito por Galileo Galilei (1610) en donde comunica sus descubrimientos con el telescopio. Foto: Guillermo A. Lemarchand tomada en la Colección de Libros Raros de la Universidad de Colorado en Boulder.

*"…Así pues, está claro que nuestra Luna es para los que estamos en la Tierra y no para los demás globos, mientras que esas cuatro lunitas que se hallan en Júpiter no son para nosotros, sino que los cuerpos que orbitan sirven respectivamente a cada uno de sus globos planetarios y a sus habitantes. De estas consideraciones concluimos que es muy grande la probabilidad de que existan habitantes en Júpiter, algo que también pensaba Tycho Brahe, basándose exclusivamente en la enormidad de ese globo."*

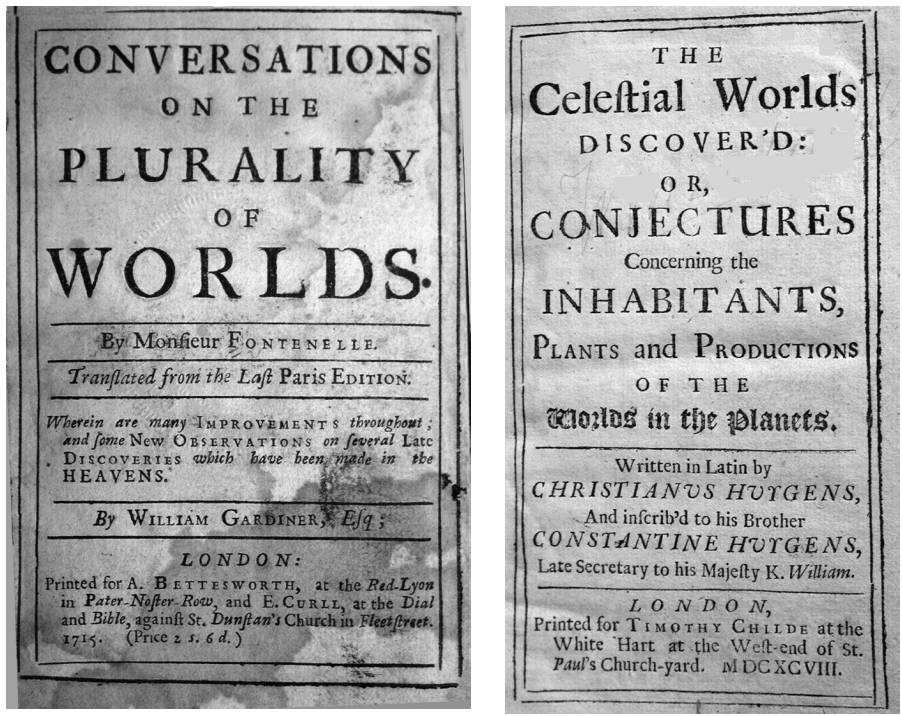

**Fig.3.** Portada de dos libros que ejercieron una poderosa influencia en la discusión sobre la pluralidad de mundos habitados: Conversaciones sobre la Pluralidad de Mundos por B. de Fontenelle (1686) y *Mundos Celestiales: Descubrimientos y conjeturas concernientes a los habitantes, plantas y producciones de los mundos en los planetas* por C. Huygens (1698). Foto: Guillermo A. Lemarchand tomada en la Colección de Libros Raros de la Universidad de Colorado en Boulder.

En los siglos sucesivos las discusiones comenzaron a incorporar argumentos cada vez más sofisticados (Lemarchand, 1992:11-36). Con el advenimiento de la *Teoría de Evolución* basada en el *principio de Selección Natural* y desarrollada independientemente por Charles Darwin (1809-1882) y Alfred R. Wallace (1823-1913), la discusión acerca de la pluralidad de mundos habitados en el universo adquiere una nueva dimensión. Si bien Darwin nunca mencionó en sus escritos la posibilidad de vida fuera de la Tierra, Wallace (1903, 1907) mantuvo una posición muy escéptica en cuanto a la posibilidad de vida extraterrestre, en particular la vida inteligente, intentando imponer una visión de universo más antropocentrista. En sus escritos utilizó el mismo tipo de línea argumental basada en la improbabilidad estadística de repetición de la serie de eventos que permitieron, a través de la selección natural, el surgimiento de la inteligencia y la civilización. Décadas después otros dos destacados biólogos evolucionistas, Ernst Mayr (1904-2005) y George G. Simpson (1902-1984) usaron argumentos similares para despreciar las posibilidades de vida inteligente en otros mundos.

Se necesitó esperar hasta mediados del siglo XX, para llevar estas pioneras discusiones filosóficas al terreno de la ciencia moderna y comenzar, por primera vez, a plantear experimentos y observaciones destinadas a encontrar las primeras evidencias de que la vida es un fenómeno de características universales.

## 2.    Una nueva ciencia en busca de identidad

La búsqueda de vida en el universo –como rama de la ciencia-  ha venido extendiendo su campo de aplicación y desarrollo durante las últimas décadas. En consecuencia, han surgido a través de los años una multiplicidad de términos para describir sus múltiples facetas de estudio. Palabras tales como *astrobiología, cosmobiología, exobiología, xenobiología* y *bioastronomía* han venido usándose indistintamente, para describir las actividades vinculadas a los estudios del origen de la vida en la Tierra y su búsqueda en el resto del universo. El alcance y aplicación de sus distintos significados se suele superponer y confundir. ¿Cuál es, entonces, el término más adecuado para describir esta actividad?  La respuesta dependerá de quién responda. Muchos de estas denominaciones tienen casi un siglo de antigüedad (Lemarchand 1998). En lo que sigue se presentará una versión actualizada de un trabajo previo (Lemarchand 2000).

El término *"astrobiología"* fue acuñado por vez primera en la literatura académica, en una serie de artículos especializados escritos por René Berthelot (1932-1937) y luego compilados en un volumen publicado originalmente en francés (Berthelot 1938).

En estos trabajos Berthelot considera que la astrobiología responde a una etapa intermedia en el desarrollo del pensamiento humano sobre el mundo, entre las sociedades de recolectores-cazadores y la ciencia moderna, por lo que ocuparía el lugar asignado a la metafísica en el esquema de Comte. En esta concepción, la astrobiología es una combinación de la creencia en las regularidades  astronómicas y en una interpretación animista o vitalista, de todos los fenómenos. En definitiva considera a la astrobiología como lo que comúnmente llamaríamos astrología. Para Berthelot el avance de la astrobiología se produjo en relación con el desarrollo de la agricultura, que ha llevado al reconocimiento de la periodicidad de las estaciones y la creación de un calendario sistemático basado en el movimiento ordenado de las estrellas.

Posiblemente, si a mediados de la década del noventa, los responsables de la NASA hubieran encontrado estas citas originales de Berthelot, jamás hubieran asociado al término astrobiología con el programa de búsqueda de vida en el universo, que desarrolló la agencia espacial de EEUU desde entonces.

La historia de la ciencia muestra la intención de definir y circunscribir constantemente las áreas emergentes de investigación. Cuando la existencia de vegetación en Marte parecía posible, el astrónomo soviético, Gavriil Adrianovich Tikhov (1875-1960) sugirió el establecimiento de la *astrobotánica*, un campo que combina la astronomía y la botánica destinado a tratar de comprender las propiedades ópticas de los ecosistemas terrestres y de la hipotética vegetación marciana.

En 1947, inauguró la "Sección de Astrobotánica de la Academia de Ciencias de la República de Kazajstán" de la ex Unión Soviética. En 1953, comienza a utilizar en sus publicaciones también el término *astrobiología* como una generalización del estudio de la biología terrestre al de los sistemas vivientes en otros mundos.

En una conferencia de la Sociedad Interplanetaria Británica realizada en 1952, el físico y filósofo de la ciencia irlandés, John D. Bernal (1901-1971) amplió sus especulaciones sobre el origen de la vida en el universo, afirmando que "la biología del futuro no estaría limitada a la Tierra", y ésta abarcaría un espectro muchísimo más amplio transformándose, en una verdadera *cosmobiología*. Este ha sido tal vez uno de los primeros términos utilizados, por un científico de fama internacional, para describir un campo que estudiaría las posibilidades de actividades biológicas y de la vida más allá de nuestro planeta.

En 1955, el astrónomo Otto Struve (1897-1963) acuñó en forma independiente la palabra *astrobiología*, con el objeto de describir el estudio amplio de la vida fuera de la Tierra. Struve también se desempeñó como el primer director del Observatorio Nacional de Radioastronomía (NRAO) de los EEUU. En noviembre de 1961 tuvo la responsabilidad de organizar, la *Conferencia de Green Bank*, destinada a determinar la posibilidad de detectar evidencias de vida inteligente en el universo y donde se presentó por primera vez la llamada *Ecuación de Drake*, destinada a determinar el posible número de civilizaciones tecnológicas en la galaxia.

Por entonces el término astrobiología comenzaba a popularizarse. Flavio A. Pereira (1956) de la Sociedad Brasilera Interplanetaria (organización her-mana de la tradicional Sociedad Interplanetaria Británica) publicó en Brasil, un libro en portugués sobre astrobiología, posiblemente el primero dedicado a la acepción moderna de esta temática. Al año siguiente, en junio de 1957, Albert G. Wilson, director del Observatorio Lowell, organizó el primer "Simposio Americano de Astrobiología".

A finales de la década del cincuenta, existían en EEUU dos grandes grupos de destacados científicos que tenían la responsabilidad de coordinar las investigaciones dentro del campo de la "vida extraterrestre". El primero de ellos era el *Panel 2 sobre Vida Extraterrestre del Comité Nacional de Bioastronáutica perteneciente a la Junta de las Fuerzas Armadas,* estaba presidido por Melvin Calvin (1911-1997) y contaba con la participación de Carl Sagan (1934-1996). El segundo grupo conformaba el *Panel sobre Vida Extraterrestre de la Academia Nacional de Ciencias* que fue presidido por el destacado biólogo y genetista Joshua Lederberg (1925-2008). Durante esos días, las temáticas eran dominadas por la búsqueda de vida en Marte y el desarrollo de los dispositivos de abordo, destinados a las primeras sondas planetarias. Esta actividad científica era –en aquella época- denominada genéricamente como *Bioastronáutica*.

Lederberg (1960) publicó un artículo seminal en el que acuñó el término *exobiología* para describir lo que llamó "la biología de origen extraterrestre". Años más tarde, también introduce el término *esobiología*, en referencia a la biología de la propia Tierra. Según Lederberg: "El objetivo primordial de la investigación exobiológica es comparar los diversos modelos de evolución química de los planetas, haciendo hincapié en aquellas características dominantes que están presentes en cada uno de ellos."

En febrero de 1963, se organizó el primer simposio internacional sobre exobiología en el ámbito del Laboratorio de Propulsión a Chorro de la NASA En el prólogo de las actas de la reunión, sus editores Mamikunian y Briggs (1965) escribieron: "La biología fuera del entorno terrestre ha sido definida como exobiología por el profesor Joshua Lederberg de la Universidad de Stanford, mientras que otros prefieren el término cosmobiología para referirse al estudio de las biologías presentes en el sistema solar, la galaxia, e incluso los sistemas extragalácticos."

En 1964, George G. Simpson, caracterizó, en forma muy filosa, el uso y abuso de la expresión *exobiología,* que le resultaba por lo menos muy curiosa, si se tiene en cuenta el hecho que esta "ciencia" todavía tenía que demostrar que su objeto de estudio existía realmente. Simpson no compartía la euforia de algunos de sus colegas exobiólogos a quienes consideraba que la actividad que desempeñaban los estaba transformando rápidamente en "ex-biólogos" (Simpson 1964).

En 1965, Gilvert Levin distintos términos empleados en el pasado para definir la búsqueda de biologías de origen extraterrestre. Finalmente llega a la conclusión que el término más adecuado para definir la actividad debía ser *xenobiología* (Levin 1965).

A finales de la década de los cincuenta, con el incremento exponencial de la sensibilidad de los equipos radioastronómicos, se propone su uso para detectar señales electromagnéticas de hipotéticas civilizaciones extraterrestres (Cocconi y Morrison, 1959). En 1960, Frank D. Drake llevó a cabo el primer experimento, utilizando las instalaciones del NRAO. Desde entonces, el objetivo de estos proyectos ha sido la detección de señales artificiales provenientes de estrellas cercanas, utilizando los telescopios de radio disponibles en todo el mundo. El descubrimiento de una señal que no pudiera ser asociada a una fuente de origen natural, estaría indicando la existencia de vida inteligente más allá de la Tierra. Durante los años sesenta, se originaron varios trabajos científicos, libros, y reuniones académicas, destinadas a desarrollar las metodologías más apropiadas de detección, que fueron por entonces categorizadas bajo el lema de "Comunicación Interestelar" (Hoerner 1961; Cameron, 1963; Kaplan 1971, Cameron y Ponnamperuma 1974).

En 1965, con el fin de organizar una reunión internacional sobre el problema de las civilizaciones extraterrestres, el profesor Rudolf Pesek acuñó el acrónimo *CETI* para definir los programas de Comunicación con Inteligencias Extra Terrestres. La elección estaba relacionada también con el hecho bien conocido que *Ceti* en inglés es el genitivo de Cetus (la familia a la cual pertenecen los delfines). Los primeros intentos de comunicación entre especies se habían hecho en la década del cincuenta con los delfines. Por otra parte, una de las estrellas que había sido observada por Frank Drake en su proyecto OZMA fue precisamente *Tau Ceti*. Por estas razones, los participantes de la Conferencia de Green Bank, se comenzaron a considerar a sí mismos como "Miembros de la Orden del Delfín".

El 15 de mayo de 1965, durante la séptima sesión de la Academia Internacional de Astronáutica (IAA), Pesek propuso "CETI" como el tema de un simposio de tres días de IAA. Después de varias reuniones y la conformación de un comité internacional de organización el programa recomendado fue el siguiente: (1) Perspectivas astronómicas de la vida; (2) Origen de la vida en la Tierra y búsqueda de vida en el Sistema Solar; (3) Evolución de la inteligencia; (4) Evolución de sociedades tecnológicas; (5) Sitios potenciales para la búsqueda de inteligencia extraterrestre; (6) Adquisición y procesamiento de las señales; (7) CETI y su impacto en la Humanidad y (8) Perspectivas y recomendaciones para futuras investigaciones. Una observación cuidadosa muestra que casi cincuenta años después, las reuniones internacionales, conservan la misma agenda.

**Fig. 4.** Como ejemplo del interés público acerca de la búsqueda de vida inteligente en el universo, la UNESCO publica un detallado artículo sobre la comunicación interestelar en un número de El Correo en enero de 1966.

Finalmente, el Comité Organizador CETI de la IAA canceló la conferencia prevista y decidió organizar una sesión de medio día "CETI: reunión de revisión durante el Congreso de la Federación Astronáutica Internacional en Viena (1972). Durante las diversas reuniones de las distintas organizaciones internacionales que tuvieron lugar entre 1965 y 1971, la temática "Comunicación Interestelar" fue sustituida por el acrónimo "CETI".

El simposio anterior fue sustituido por una conferencia soviético-norteamericana sobre CETI celebrada en Byurakan entre el 5-11 de septiembre de 1971. La misma se organizó en conmemoración al décimo aniversario de la Conferencia de Green Bank. La reunión se celebró a la vista del monte Ararat, en el Observatorio Astrofísico Byurakan (en la ex República Soviética de Armenia). Fue, patrocinado por las Academias de Ciencias de los Estados Unidos y de la Unión Soviética. Así como en Green Bank, el principio organizador de la conferencia fue la *Ecuación de Drake* (Sagan, 1973). En una forma u otra, por estos días se consideraba a la Ecuación de Drake como la base epistemológica de la investigación CETI.

Después de una serie de talleres organizados por la NASA, a mediados de los años setenta, algunos expertos temieron que un hipotético mensaje extraterrestre de una sociedad avanzada podría hacer perder la fe en la capacidad de la raza humana y eventualmente privar de la iniciativa para hacer nuevos descubrimientos, o generar consecuencias negativas para la humanidad (Morrison et al 1977). Pese a que si se recibiera un mensaje no existiría ninguna obligación de responder, los organizadores de estos talleres decidieron cambiar la sigla de comunicación con inteligencias extraterrestres (CETI) a la de Búsqueda de Inteligencias Extra Terrestres (SETI). Desde entonces, este término se viene utilizando ampliamente.

En los años ochenta, con el fin de obtener algún apoyo institucional de las autoridades de la NASA y del Congreso de EEUU, el proyecto SETI fue rebautizado con el nombre SETI MOP (Proyecto de observación en microondas). Una década después, por razones similares, la sigla se volvió a reemplazar por HRMS (Relevamiento de Microondas en Alta Resolución). Con este cambió se eliminó definitivamente la palabra SETI. El proyecto HRMS de la NASA involucraba un programa de observación de diez años y una inversión que superaba ampliamente los cien millones de dólares. Comenzó a funcionar el 12 de octubre de 1992, conmemorando el quinto centenario del descubrimiento de América. Finalmente, en 1993, el Congreso de EEUU canceló toda participación de la NASA en cualquier programa SETI.

La sigla SETI puede introducir cierta confusión al lego ya que –por el momento- no se dispone de ninguna tecnología que permita detectar "inteligencias" en forma directa a través de distancias interestelares. Lo único que se puede hacer, es intentar detectar manifestaciones de actividades tecnológicas. Si en el medio del océano se encontrar una flotando una botella con un mensaje en su interior es posible que no pueda entender el significado del mensaje, pero con toda certeza se podrá inferir que un ser inteligente fabricó la botella y colocó un mensaje en ella.

Popper consideraba que la principal actividad científica consiste en falsar teorías. Por lo tanto, con el fin de poner a prueba la hipótesis original de que existe vida inteligente en el universo, la estrategia que se está desarrollando es diseñar una búsqueda exhaustiva de Actividades Tecnológicas de origen

Extraterrestre. Lemarchand (1992, 1994) propuso utilizar la sigla en inglés SETTA *(Search for Extra Terrestrial Technological Activities)* para definir en forma más precisa el tipo de investigación que se lleva a cabo.

La participación activa de la Unión Astronómica Internacional (UAI) en la búsqueda de vida extraterrestre fue iniciada con el patrocinio de la reunión conjunta de varias comisiones durante la XVII Asamblea General de Montreal en 1979. Allí se organizó un simposio denominado: "Estrategias para la Búsqueda de Vida en el Universo" (Papagiannis 1980). Esta reunión terminó con una sesión abierta celebrada en el gran auditorio de la Universidad de Montreal. Allí asistieron más de 1000 astrónomos y parte del auditorio siguió las presentaciones de pie.

Tras el éxito de la reunión de Montreal, en 1982, durante XVIII Asamblea General de la Unión Astronómica Internacional (IAU) realizada en Patras, Grecia se creó la Comisión 51. La misma recibió la denominación de *Bioastronomía* (búsqueda astronómica de la vida –bios- en el universo) y se estableció para consolidar el esfuerzo internacional en la búsqueda de vida extraterrestre. Desde entonces, la Comisión 51 ha organizado 9 conferencias internacionales sobre *Bioastronomía* (ver detalles en Lemarchand 2000).

Junto con la IAU, la Sociedad Internacional para los Estudios sobre el Origen de la Vida (ISSOL) ha desempeñado un papel fundamental en la promoción y articulación de los trabajos vinculados al origen de la vida en la Tierra y su posible existencia en otros mundos. Cada tres años organiza conferencias internacionales con una creciente participación de representantes de Iberoamérica.

Después del anuncio del descubrimiento del meteorito ALH80001, en 1996, donde se especulaba acerca de la posibilidad de bacterias de origen marciano, la NASA puso en marcha un ambicioso Programa de Astrobiología. Este inyectó decenas de millones de dólares al sistema de investigación y desarrollo, logrando que en pocos años, la comunidad de científicos que se dedicaban a la exobilogía, bioastronomía, etc. se multiplicaran varios órdenes de magnitud.

En su acepción moderna la *astrobiología* es definida como el estudio de los orígenes, evolución, distribución y futuro de la vida en el universo. Para su desarrollo se requieren conceptos fundamentales acerca de la vida y entornos habitables que ayudan a reconocer biosferas que pudieran ser muy diferentes de las nuestras. La astrobiología abarca la búsqueda de planetas potencialmente habitados más allá de nuestro Sistema Solar, la exploración de Marte y los planetas exteriores, y las investigaciones de laboratorio sobre los orígenes y la evolución temprana de la vida, así como estudios sobre el potencial de la vida para adaptarse a los retos del futuro, tanto en la Tierra como en el espacio. Es una investigación netamente interdisciplinaria y para su desarrollo es necesario combinar los saberes de la biología molecular, la ecología, la ciencia planetaria, astronomía, ciencias de la información, tecnologías de exploración espacial, y otras disciplinas afines. La astrobiología promueve una comprensión más amplia e integradora de los fenómenos biológicos, planetarios y cósmicos.

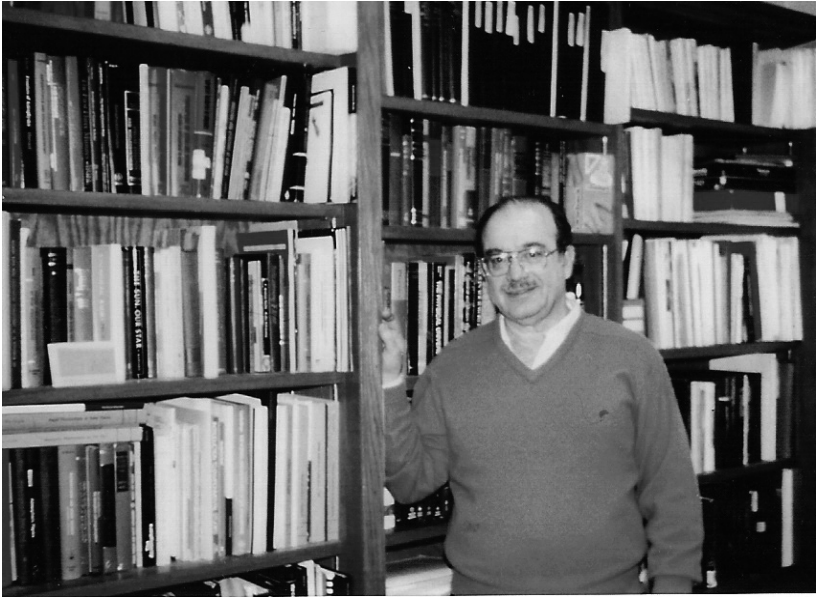

**Fig. 5.** Prof. Michael D. Papagiannis (1932-1998), director del Departamento de Astronomía de la Universidad de Boston y fundador de la Comisión 51 de Bioastronomía de la Unión Astronómica Internacional. Foto: Guillermo A. Lemarchand (c. 1988).

## 3. Algunos hitos en el desarrollo de la astrobiología en Iberoamérica

Durante el 2010 se está celebrando, en diversos países de América Latina, el bicentenario de su independencia. Es curioso que algunos de los patriotas que participaron en dicho proceso fueran también los primeros en especular –en estas latitudes- acerca de la vida en otros mundos. Uno de ellos fue Manuel

Moreno (1782-1857), político y médico argentino, hermano menor de Mariano Moreno, prócer de la Revolución de Mayo.

Manuel Moreno apoyó la Revolución de Mayo y ejerció cargos menores dependientes de la Primera Junta. En 1811 acompañó en su misión diplomática a Gran Bretaña a Mariano, pero éste murió en alta mar por una intoxicación con un medicamento que le administró mal el capitán del barco. Al llegar a Londres permaneció sin una misión clara, aprovechando el tiempo en estudiar y escribir su *Vida y Memorias de Mariano Moreno*. Regresó a Buenos Aires en 1812 y fue nombrado secretario del Segundo Triunvirato, que estaba dominado por la Logia Lautaro, a la que se incorporó. Durante las sesiones de la Asamblea del Año XIII, defendió firmemente desde la prensa la forma republicana de gobierno.

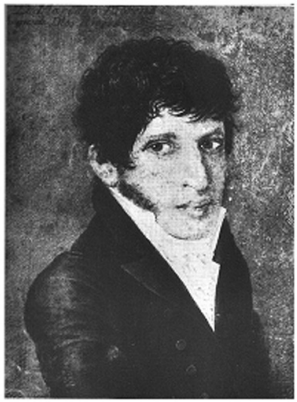
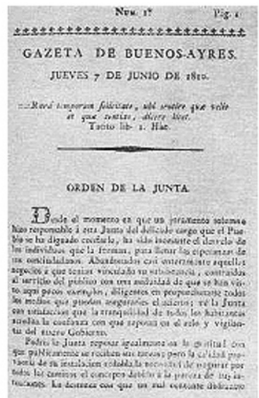

**Fig. 6.** Manuel Moreno (1782-1857), político y médico argentino, hermano del prócer de la Revolución de Mayo (1810), Mariano Moreno (1778-1811). Fundó el Departamento de Medicina de la Universidad de Buenos Aires y en sus textos en la Gazeta de Buenos Ayres especuló acerca del origen de la vida en la Tierra y en otros mundos.

Años más tarde fundó el Departamento de Medicina de la Universidad de Buenos Aires. Desde 1823, dictó la cátedra de química de la universidad, siendo el primero en dar clases públicas de esa disciplina en el país, hecho que le valió el mote de "Don Óxido". En sus escritos en *La Gazeta* discutió acerca del origen de la vida en la Tierra y de la posible existencia de vida en otros mundos.

Otro destacado patriota rioplatense que especuló acerca de la vida en otros mundos fue Vicente Fidel López (1815-1903). En cartas particulares comentando la obra *Cosmos* de Alexander von Humboldt (c. 1847-1850) discute el origen del vulcanismo y acerca de la posibilidad que éste último esté presente en otros mundos como fuente de generación de vida.

Es bien conocido que parte del trabajo de campo de Charles Darwin que le permitió –como naturalista- concebir y desarrollar la teoría del origen de las especies fue realizado en América del Sur. En su recorrido con el Beagle, tuvo la oportunidad de recorrer las pampas argentinas, la cordillera de los Andes, el desierto de Atacama y las islas Galápagos.

En agosto de 1924, Marte tuvo un acercamiento orbital a la Tierra. David Tood, profesor de astronomía del Armhest College de EEUU, propuso –por entonces- detectar señales radiofónicas emitidas por hipotéticos habitantes marcianos. Dichas señales podrían estar siendo dirigidas hacia la Tierra aprovechando la coincidencia de que en ese momento la distancia entre ambos planetas sería mínima. Todd inició una campaña internacional para que los grandes transmisores de radio, instalados en todo el mundo, cesaran de transmitir durante 5 minutos, en cada hora, para apuntar todos los receptores hacia Marte sin la interferencia de las señales de origen terrestre. El 21 de agosto de 1924, el periódico *The New York Times,* informó que el Prof. Todd había tenido una conversación con el Embajador Pueyrredon, representante argentino en Washington, y que la Argentina se había comprometido a que sus más poderosos transmisores hicieran silencio de radio de acuerdo al esquema presentado. Aparentemente, Brasil se habría sumado también a dicha iniciativa. Este es el antecedente histórico más antiguo de un verdadero proyecto SETI de carácter internacional y de la participación de América Latina en el mismo.

El gran precursor de la ciencia biológica en México, Alfonso L. Herrera (1868-1944) fue reconocido por sus grandes méritos en múltiples instituciones científicas mundiales, ganador de las Palmas Académicas de Francia, y el único mexicano miembro de la *Academia del Lincei* de Roma al lado de otros científicos como Einstein, Newton y Ohm. Herrera desempeñó un papel fundamental en introducir nuevas teorías acerca del origen de la vida. Desarrolló la teoría de la plasmogenia para explicar el origen de la vida; en ella destaca los procesos de formación del plotoplasma, compuesto fundamental para que se manifestaran las primeras formas de vida (Herrera 1924, 1932). En sus investigaciones también demostró la síntesis abiótica de compuestos orgánicos; sin embargo no llegó a definir completamente los límites entre la materia viva y la materia inanimada. Fue el primer biólogo de la región en publicar un artículo sobre el origen de la vida en la prestigiosa revista *Science* (Herrera 1942).

Sin embargo, el científico iberoamericano que se consagró como experto mundial en temas de origen de la vida y "exobiología" fue Joan Oró (1923-2004). En 1956, fundó el Departamento de Ciencias Biofísicas en la Universidad de Houston, dónde estudió el metabolismo del ácido fórmico en los tejidos animales, investigaciones que serían clave para el estudio sobre el origen de la vida y la interpretación de la ausencia de vida en el planeta Marte. Se lo considera como uno de los fundadores de la cosmoquímica orgánica. Alcanzó fama mundial cuando en la Navidad de 1959, encerrado en su laboratorio, descubrió la síntesis de la adenina, una de las moléculas más importantes para la vida. La paradoja de su descubrimiento fue que esta sustancia pudo ser sistetizada a partir del ácido cianhídrico, uno de los productos más venenosos (Oró y Kimball 1961). Dos años más tarde publicó su artículo seminal acerca del papel que tuvieron los cometas en el origen de la vida en la Tierra (Oró 1961). En 1963, participó del Primer Simposio Internacional de Exobiología organizado en el Laboratorio de Propulsión a Chorro de la NASA (Mamikunian y Briggs 1965). Desde entonces desarrolló varios programas de investigación espacial de la NASA, como en el proyecto Apolo para el análisis de las rocas y otras muestras de material de la Luna, y en el proyecto Viking, encargándose del desarrollo de un instrumento para el análisis molecular de la atmósfera y la materia de la superficie del planeta Marte. En 1999, junto a Julián Chela Flores y el autor dirigieron la *Primera Escuela Iberoamericana de Astrobiología* (Chela Flores et al 2000).

En 1966 se funda el Instituto Argentino de Radioastronomía (IAR), su primer director fue Carlos María Varsavsky (1933-1983) un astrofísico de la Universidad de Harvard. En 1968 publica el primer libro de un científico latinoamericano sobre búsqueda de vida inteligente en el universo (Varsavsky, 1968). Escrito en un estilo ameno, el libro, ejerció una poderosa influencia en varias generaciones de jóvenes científicos. Dentro del grupo fundador del IAR se encontraban también dos jóvenes radioastrónomos que desempeñarían un papel fundamental en el posterior desarrollo de la búsqueda de vida extraterrestre en la región: Fernando Raúl Colomb (1939-2008) y Valentín Boariakoff (1938-1999).

A principios de la década del setenta, el Prof. Raúl Ondarza, de la Universidad Nacional Autónoma de México, organizó un simposio sobre origen de la vida en donde participaron, entre otros, figuras de la talla de Stanley Miller,

Joan Oró, Carl Sagan, Cyril Ponnamperuma. Allí, Alicia Negrón Mendoza, conoció a Ponnamperuma y se fue a trabajar con él, obteniendo el primer doctorado de América Latina en un tema vinculado al origen de la vida.

En 1975, Antonio Lazcano-Araujo, Alfredo Barrera y Juan Luis Cifuentes, llevaron a Aleksandr I. Oparin (1894-1980) a México y organizaron un simposio internacional en su honor (Lazcano y Barreda 1978).

Para 1977, Antonio Lazcano ya tenía su curso universitario de "Origen de la vida", había publicado un texto que fue traducido a varios idiomas (Lazcano, 1977) y se encontraba trabajando con Stanley Miller y Joan Oró. Sin duda, a partir de entonces "Toño" se transformó en un pilar fundamental en el cual se apoyaron los estudios sobre origen de la vida en México. El trabajo de "Toño" promovió la formación de decenas de doctorandos y expertos internacionales de origen mexicano en esta temática. Merced al reconocimiento internacional logrado, se desempeñó, en dos oportunidades, como Presidente de la Sociedad Internacional sobre Estudios del Origen de la Vida (ISSOL).

En 1977 el ingeniero argentino, Valentín Boriakoff, uno de los fundadores del IAR, resuelve la manera de registrar imágenes en formato de 33 ½ r.p.m. para el disco interestelar de las naves Voyager I y II. Colaborando de esta manera en el diseño de un mensaje interestelar destinado a una hipotética civilización extraterrestre. Una foto del propio Boriakoff fue incluida dentro del contenido de los discos de las naves *Voyager*, imagen que se perpetuará en los confines del medio interestelar, por al menos, unos doscientos mil años…

En la primavera de 1979, un grupo de profesores de la Facultad de Ciencias Físicas y Matemáticas de la Universidad de Chile, organizaron un curso sobre el enfoque científico de la búsqueda de vida inteligente en el universo que se desarrolló en Santiago de Chile (Campusano 1985). Posiblemente este fue el primer curso formal de carácter universitario sobre vida en el universo dictado en la región.

En junio de 1984, la recién formada Comisión 51 de Bioastronomía de la Unión Astronómica Internacional (IAU) organizó en la ciudad de Boston el primer simposio oficial de la IAU dedicado a los estudios de búsqueda de vida en el universo. El astrónomo nacido en Uruguay y nacionalizado argentino, Félix Mirabel participó en dicha reunión y promovió la posibilidad de comenzar a realizar investigaciones radioastronómicas de búsqueda de señales de origen extraterrestre desde el hemisferio sur (Mirabel, 1984). Años más tarde, "Raúl" Colomb sería designado como presidente de la Comisión 51 de la IAU.

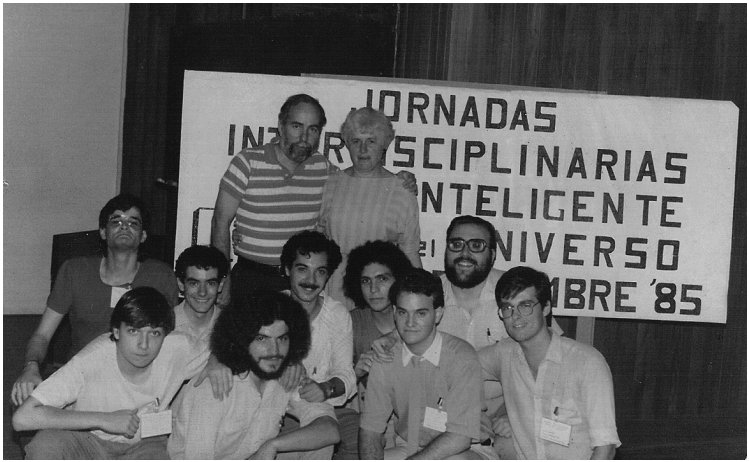

**Fig. 7.** Grupo de jóvenes estudiantes de física junto a R. Bruce Crow, especialista en SETI de la NASA, durante la organización de las Jornadas Interdisciplinarias sobre la Vida Inteligente en el Universo, Facultad de Ciencias Exactas y Naturales de la Universidad de Buenos Aires, diciembre de 1985. Foto: G. A. Lemarchand. (c.1985).

En diciembre de 1985 un grupo de entusiastas estudiantes de física de la Facultad de Ciencias Exactas y Naturales (FCEN) de la Universidad de Buenos Aires, organizaron las *Primeras Jornadas Interdisciplinarias sobre Vida Inteligente en el Universo* (Lemarchand, 1986). El evento se desarrolló durante tres días consecutivos y contó con la participación de más de una veintena de especialistas, entre los que se destacaban Robert Bruce Crow (especialista en temas de SETI de la NASA); Félix Mirabel (radioastrónomo del Observatorio de Arecibo en Puerto rico), Miguel Ángel Virasoro (físico de la Universidad de Roma y más tarde director del ICTP, Centro de Física Teórica de Trieste), Fernando R. Colomb (Director del IAR); Gregorio Klimovsky (epistemólogo y decano de la FCEN), Aldo Armando Cocca (jurista de renombre internacional, especialista en derecho del espacio). Durante la reunión participaron cerca de 500 estudiantes y público general que colmaron el aula magna de la FCEN.

Las "jornadas" facilitaron la firma de un convenio de cooperación entre el Instituto Argentino de Radioastronomía y *The Planetary Society* para la consolidación de un programa observacional de SETI desde el hemisferio sur.

El 7 de octubre de 1986 comenzaron las primeras observaciones SETI desde el IAR (Colomb y Lemarchand 1989). El grupo de investigación liderado por "Raúl" Colomb estaba integrado por María Cristina Martín y el autor. Un año después se había transformado en el proyecto más importante realizado, hasta entonces, en el hemisferio sur (Lemarchand 1992; Colomb et al 1992).

Mientras las antenas del IAR se encontraban rastreando señales artificiales de origen extraterrestre alrededor de las estrellas cercanas del hemisferio sur, en 1987 se organizó en Buenos Aires un seminario para discutir los aspectos radiastronómicos, sociales y jurídicos de la detección de señales electromagnéticas de origen extraterrestre (Cocca 1988). La reunión estuvo coordinado por el destacado jurista internacional Aldo Armando Cocca, uno de los autores del Tratado del Espacio (1967) y de la Luna y otros Objetos Celestes (1979). Su objetivo fue discutir los alcances de un protocolo internacional que se estaba elaborando dentro del ámbito del Instituto Internacional de Derecho Espacial (IISL), la Academia Internacional de Astronáutica (IAA), la Unión Astronómica Internacional (IAU) y la Federación Internacional de Astronáutica (IFA),

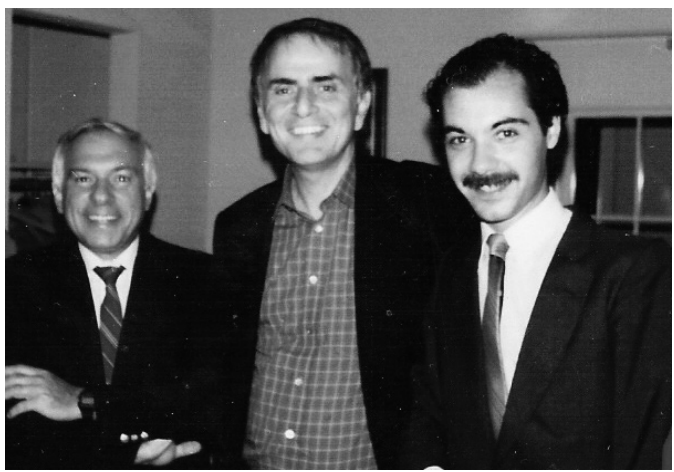

**Fig. 8.** De izquierda a derecha: Y. Terzian, Director del Departamento de Astronomía de la Universidad de Cornell; Carl Sagan, Director del Laboratorio de Estudios Planetarios de la Universidad de Cornell y Presidente de The Planetary Society y Guillermo A. Lemarchand (c. 1988).

En enero de 1988 el autor visitó el programa SETI que estaba funcionando desde la Universidad de Harvard y estableció las bases para la redacción de un convenio que redactó unas semanas después en la sede de *The Planetary Society* (TPS) en Pasadena, California. En octubre de 1988, durante una conferencia internacional sobre búsqueda de inteligencias extraterrestres organizada por TPS en el Centro de Ciencias de Ontario (Canadá), su presidente, Carl Sagan y el director del IAR, Fernando R. Colomb, anunciaron la firma de un convenio de cooperación entre el Consejo Nacional de Investigaciones Científicas y Tecnológicas (CONICET) de Argentina y *The Planetary Society*, para la construcción de una réplica de un analizador espectral de 8,4 millones de canales, con una resolución espectral de 0,05 Hertz por canal, que se encontraba en operación en el radiotelescopio de 26 m de la Universidad de Harvard.

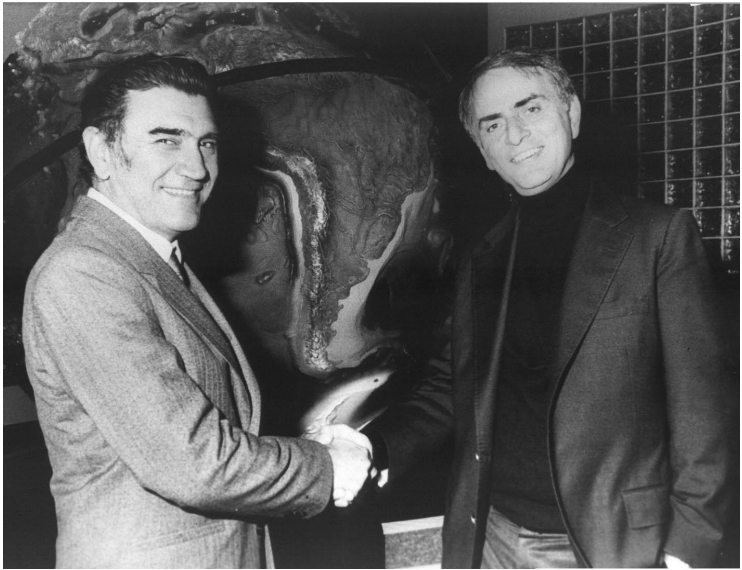

**Fig. 9.** Fernando R. Colomb y Carl Sagan durante la firma del Convenio entre The Planetary Society y el Instituto Argentino de Radioastronomía en el Museo de Ciencias de Ontario en Canadá en octubre de 1988. Foto: Guillermo A. Lemarchand.

Dos ingenieros argentinos, Eduardo Hurrell y Juan Carlos Olalde, permanecieron por un año en la Universidad de Harvard, bajo la dirección del Prof. Paul Horowitz, construyendo el analizador espectral denominado META II (Megachannel Extra Terrestrial Assay).

Finalmente, a las 10 de la mañana de un soleado 12 de octubre de 1990, el proyecto META II era inaugurado en la Argentina por Louis Friedman (director ejecutivo de TPS) y Fernando R. Colomb, quienes apuntaron la antena II del IAR hacia la constelación de la Cruz del Sur. A partir de entonces, por primera vez en la historia de la humanidad, se disponía de un programa de búsqueda de señales de origen extraterrestre que observaba constantemente todo el cielo accesible desde la Tierra. El hemisferio norte era relevado desde el radiotelescopio de la Universidad de Harvard en Oak Ridge y el hemisferio sur desde el IAR. El sistema argentino observó sistemáticamente todo el cielo del hemisferio sur, entre declinaciones de -10º y -90º, en diversas frecuencias durante más de una década (Lemarchand et al. 1997).

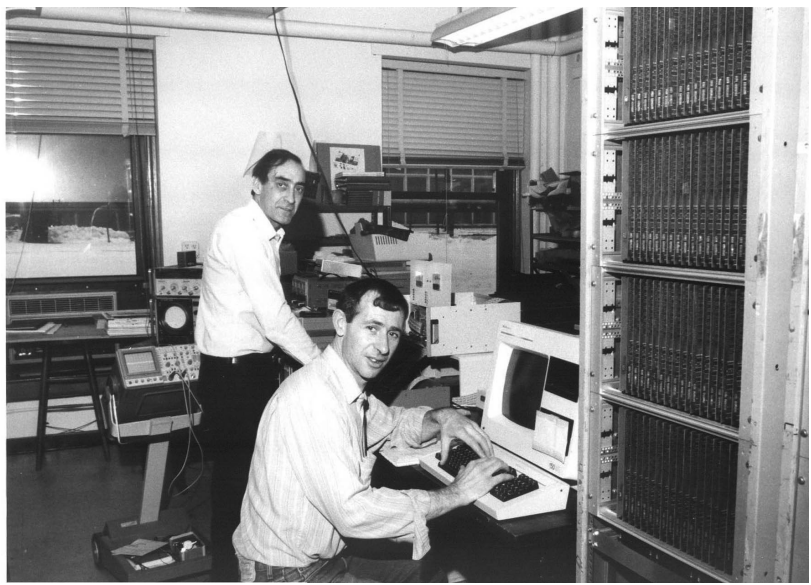

**Fig. 10.** Los ingenieros argentinos Juan Carlos Olalde y E. Eduardo Hurrell en las instalaciones del laboratorio Lyman en la Universidad de Harvard, poniendo a punto el analizador espectral de 8,4 millones de canales destinado al IAR bajo la dirección del Prof. Paul Horowitz.

Mientras tanto, muy cerca de allí, en la Facultad de Ciencias de la Universidad de la República, en Montevideo, Uruguay, Julio A. Fernández un especialista internacional en cometas, junto con un nutrido conjunto de expertos internacionales en diversas disciplinas, organizó en marzo de 1988, un ciclo de conferencias *"Vida y Cosmos: Un enfoque interdisciplinario"* (Fernández 1988) y luego en febrero de 1995 la una escuela de posgrado *"Vida y Cosmos: Nuevas reflexiones"* (Fernández y Mizraji 1995). Ambos eventos generaron una gran efervescencia entre estudiantes y docentes y sirvieron de modelo para la organización de la que sería, posteriormente, la Escuela Iberoamericana de Astrobiología.

Entre 1992 y 1994, el físico venezolano Julián Chela Flores, comenzó a trabajar junto a Cyril Ponnamperuma en el ICTP, allí organizaron tres conferencias internacionales sobre evolución química y exobiología. Luego del fallecimiento de Ponnameruma, Julián Chela Flores, continuó con la tradición iniciada y organizó, entre 1995 y 2003, cuatro conferencias internacionales más. Chela-Flores desempeñó un papel fundamental en promover la participación de jóvenes científicos de América Latina junto a destacadísimas figuras en el campo de la exobiología y astrobiología internacional. En 1999, co-dirigió junto a Joan Oró y el autor, la Primera Escuela Iberoamericana de Astrobiología (Chela Flores et al. 2000).

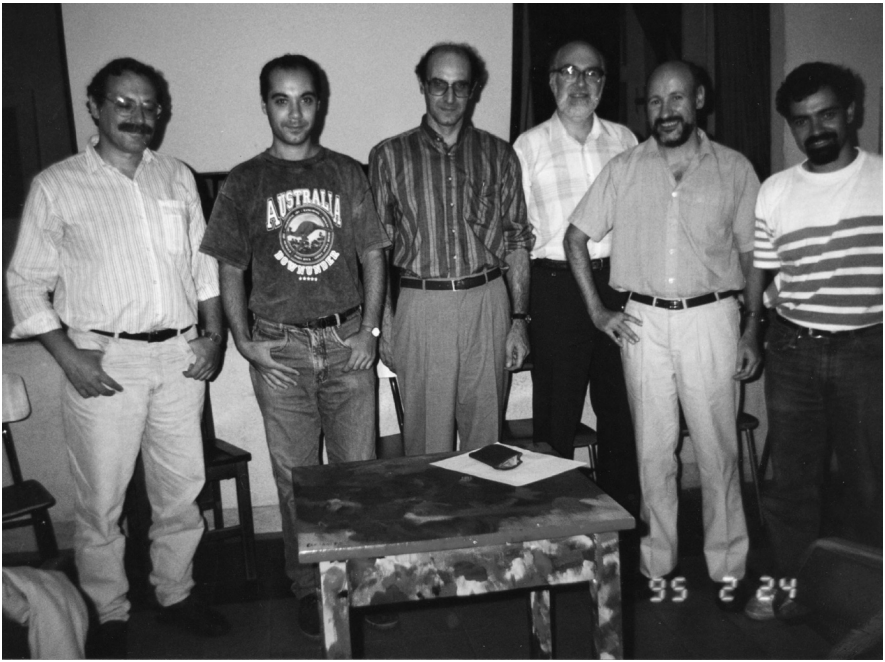

**Fig. 11.** Un conjunto de profesores de la Escuela Vida y Cosmos II (Montevideo, 1995). De izq. a der. Ricardo Erlich, actual Ministro de Educación y Cultura de Uruguay, Guillermo A. Lemarchand, Eduardo Mizraji, Luis Elbert, Julio A. Fernández y Gonzalo Tancredi. Foto: Guillermo A. Lemarchand.

En 1998 se crea en España el Centro de Astrobiología de Madrid (CAB). Su origen se remonta a la propuesta presentada a la NASA por un grupo de científicos españoles y norteamericanos liderados por Juan Pérez-Mercader para unirse al entonces recién creado NASA Astrobiology Institute (NAI). Después de un minucioso análisis y evaluación de la propuesta y tras un intercambio de cartas a nivel de Gobierno, el CAB fue integrado en el NAI en abril de 2000, convirtiéndose de esta manera en el primer Miembro Asociado al NAI fuera de Estados Unidos. El otro centro Miembro Asociado del NAI es, desde 2003, el Australian Center for Astrobiology.

El CAB fue creado como centro mixto entre el Consejo Superior de Investigaciones Científicas (CSIC) y el Instituto Nacional de Tecnología Aeroespacial (INTA), y con el apoyo de la Comunidad Autónoma de Madrid (CAM). El entonces presidente del INTA y Secretario de Estado de Defensa, Pedro Morenés y el presidente del CSIC, César Nombela, firmaron el 19 de noviembre de 1999 el acuerdo de constitución sobre la base del acuerdo de cooperación entre ambas instituciones de 1991 siendo director general del INTA el profesor Emilio Varela. Su objetivo inicial fue establecer un entorno investigador verdaderamente transdisciplinar para el desarrollo de la nueva ciencia de la astrobiología, con una nueva y específica contribución de una metodología común basada en las teorías de complejidad y en la aplicación del método científico a la Vida.

**Fig.12.** Afiche de convocatoria de para la Primera Escuela Iberoamericana de Astrobiología (Caracas, 1999)

Durante una reunión celebrada en el Centro de Física Teórica de Trieste en 1997, bajo el liderazgo de Julián Chela Flores y con el apoyo de Joan Oró y el autor, se comenzó a planificar la organización de la Primera Escuela Iberoamericana de Posgrado en Astrobiología. La misma se desarrolló en el Instituto de Estudios Avanzados, en el predio de la Universidad Simón Bolívar, en la ciudad de Caracas (Venezuela), en noviembre de 1999. La Escuela estuvo financiada por el ICTP, la UNESCO, el Centro Internacional de Ingeniería Genética y Biotecnología de Trieste, la NASA, la Universidad Simón Bolívar, la Agencia Espacial Europea, el SETI Institute, The Planetary Society, el Centro Latinoamericano de Física, la Red Latinoamericana de Biología, la Academia de Ciencias de América Latina, las Fundaciones Joan Oró y Alberto Vollmer y el Colegio Emil Friedman. La escuela contó con la participación de una veintena de profesores, en su mayoría, de origen iberoamericano y el idioma oficial fue el español. Este hecho generó una altísima sinergia y participación entre los estudiantes y profesores (Chela Flores et al 2000).

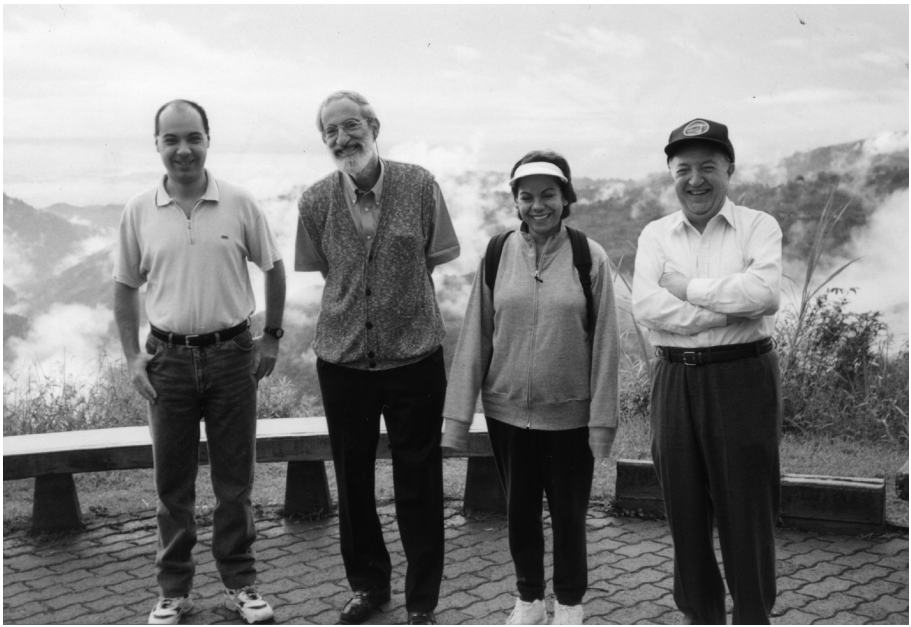

**Fig. 13.** Co-directores de la Primera Escuela Iberoamericana de Astrobiología. De izq. a der.: Guillermo A. Lemarchand, Julián Chela-Florez, una estudiante, Joan Oró. Foto: G.A. Lemarchand (Caracas, 1999).

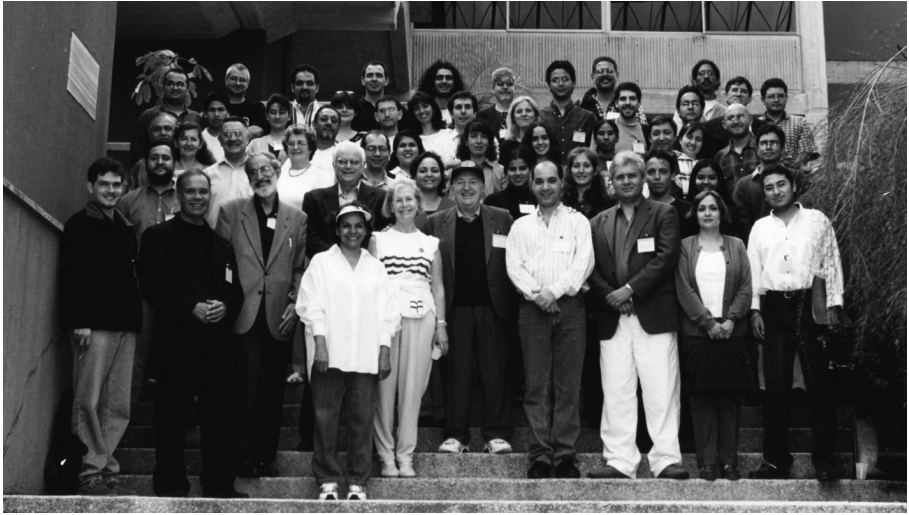

**Fig.14.** Participantes de la Primera Escuela Iberoamericana de Astrobiología (Caracas, 1999).

Durante la reunión se propuso la posibilidad de crear una Red Iberoamericana de Astrobiología para promover los estudios en la región, la elaboración de un libro de texto universitario en español acerca de la temática de vida en el universo y otras acciones similares tendientes al fortalecimiento de la cooperación Sur-Sur en esta floreciente rama de la ciencia.

Originalmente, se había convenido organizar la Segunda Escuela de Astrobiología en la ciudad de Oaxaca (México) en el año 2002, en coincidencia con la reunión trienal de ISSOL. Sin embargo, problemas organizativos impidieron concretar dicha iniciativa.

En el año 2006, merced al emprendimiento de Gustavo Porto de Mello, se organizó en la Universidad de Río de Janeiro, un taller sobre Astrobiología que contó con la participación de un centenar de científicos y estudiantes de varias ramas de la ciencia. A principios del 2008, en la Universidad de San Pablo se organizó también un exitoso seminario internacional acerca del Origen de la Vida en la Tierra y en el Universo.

Rápidamente en Chile y Colombia comenzaron a surgir iniciativas similares y aparecieron los primeros grupos de investigación en temas vinculados a la astrobiología.

Durante el año 2009, se celebraron una serie de eventos muy significativos para la astrobiología: Año Internacional de la Astronomía 2009; cuarto centenario de las primeras observaciones telescópicas de Galileo Galilei y de la publicación de las leyes de Kepler en su libro *Astronomía Nova*; bicentenario de la publicación de la primera teoría de evolución de J. B. Lamarck y del nacimiento de Charles Darwin; sesquicentenario de la publicación del "*Origen de las Especies*", 60 aniversario de la Oficina Regional de Ciencia de la UNESCO para América Latina y el Caribe; el cincuentenario de la publicación del primer artículo científico de SETI por G. Cocconi y P. Morrison y del artículo sobre síntesis de moléculas orgánicas en la Tierra primitiva, por Stanley Miller y Harold Urey; y los 10 años de la Primera Escuela Iberoamericana de Astrobiología.

En este contexto, la Oficina Regional de Ciencia de la UNESCO para América Latina y el Caribe en cooperación con la Facultad de Ciencias de la Universidad de la República, organizaron la *Segunda Escuela Iberoamericana de Astrobiología: Del Big Bang a las Civilizaciones* en la ciudad de Montevideo entre el 7 y 12 de septiembre de 2009. El evento fue patrocinado por la Organización de Estados Americanos (OEA), el Centro de Física Teórica Abdous Salam de Trieste (ICTP), la Academia de Ciencias de los Países en Desarrollo (TWAS), el Instituto Argentino de Radioastronomía (IAR), el Programa de Desarrollo de Ciencias Básicas (PEDECIBA) y la Dirección de Innovación , Ciencia y Tecnología para el Desarrollo (DICYT) del Uruguay.

En la misma, participaron alrededor de 70 estudiantes graduados, representando a 16 países de América Latina y el Caribe y una destacada planta de profesores de renombre internacional. Incluso se contó con una clase magistral del pionero Frank Drake, a través de una teleconferencia simultánea con California.

Al igual que lo sucedido durante la primera escuela, se volvió a repetir una gran sinergia y participación entre todos los estudiantes y profesores. Se organizaron conferencias públicas y un taller de maestros y profesores de ciencias del MERCOSUR, que llegó a contar con la participación de más de 200 profesores.

Finalmente, se convino que la Tercera Escuela Iberoamericana de Astrobiología fuera organizada en la ciudad de Barcelona, en el verano boreal de 2011. Su organización estará bajo la tutela del Prof. Jordi Gutiérrez.

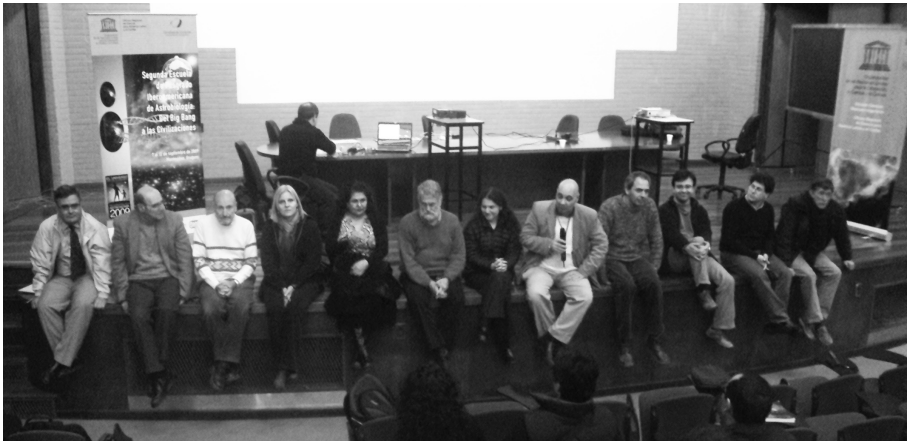

**Fig.15.** Los profesores durante la sesión de clausura de la Segunda Escuela Iberoamericana de Astrobiología (Montevideo, 2009). De izq. a der.: Álvaro Giménez, Eduardo Mizraji, Julio A. Fernández, Alicia Massarini, Antígona Segura, Ricardo Amils-Pibernat, Felisa Wolf-Simon, Guillermo A. Lemarchand, Jordi Gutiérrez, César Bertucci, Marcelo Guzmán, Antonio Lazcano, atrás de espalda, Gonzalo Tancredi. No se encuentran en la imagen, Martín Makler y Gustavo Porto de Mello.

## 4.     Epílogo

Los seguidores contemporáneos de las ideas de Epicuro, mantienen en común no solo su convicción de que no estamos solos en el universo, sino también una compresión muy fina acerca del maravilloso fenómeno de la vida y de la necesidad de garantizar su continuidad. Los Epicureanos modernos somos conscientes que es imprescindible hacer que el último factor de la Ecuación de Drake sea lo más longevo posible. Parafraseando a Joan Oró, debemos reflexionar sobre tres principios éticos que se derivan directamente de un mejor conocimiento del cosmos y de la comprensión de los mecanismos que originaron la vida en la Tierra:

*Humildad:* La vida proviene de simples moléculas

*Fraternidad:* Toda vida en la Tierra, y el *Homo sapiens* como parte de ella, tiene un origen genético común, que nos demanda solidaridad.

*Cooperación:* Necesitamos compartir los recursos limitados de nuestro planeta, en una forma sostenible para garantizar la continuidad de la vida en nuestro mundo.

## Referencias